# High-order harmonic generation in Xe, Kr, and Ar driven by a 2.1-μm source: high-order harmonic spectroscopy under macroscopic effects

**Kyung-Han Hong,**[1,*] **Chien-Jen Lai,**[1] **Vasileios-Marios Gkortsas,**[1] **Shu-Wei Huang,**[1] **Jeffrey Moses,**[1] **Eduardo Granados,**[1,2] **Siddharth Bhardwaj,**[1] **and Franz X. Kärtner**[1,3]

[1]*Department of Electrical Engineering and Computer Science and Research Laboratory of Electronics,*
*Massachusetts Institute of Technology (MIT), Cambridge, Massachusetts 02139, USA,*
[2]*IKERBASQUE, Basque Foundation for Science, Spain*
[3]*DESY-Center for Free-Electron Laser Science and Department of Physics, Hamburg University, Hamburg, Germany*
[*] *kyunghan@mit.edu*

**Abstract:**

We experimentally and numerically study the atomic response and pulse propagation effects of high-order harmonics generated in Xe, Kr, and Ar driven by a 2.1-μm infrared femtosecond light source. The light source is an optical parametric chirped-pulse amplifier, and a modified strong-field approximation and 3-dimensional pulse propagation code are used for the numerical simulations. The extended cutoff in the long-wavelength driven high-harmonic generation has revealed the spectral shaping of high-order harmonics due to the atomic structure (or photo-recombination cross-section) and the macroscopic effects, which are the main factors of determining the conversion efficiency besides the driving wavelength. Using precise numerical simulations to determine the macroscopic electron wavepacket, we are able to extract the photo-recombination cross-sections from experimental high-order harmonic spectra in the presence of macroscopic effects. We have experimentally observed that the macroscopic effects shift the observed Cooper minimum of Kr from 80 eV to 60-70 eV and wash out the Cooper minimum of Ar. Measured high-harmonic conversion efficiencies per harmonic near the cutoff are ~$10^{-9}$ for all three gases.

## I. Introduction

The high-order harmonic generation (HHG) process can be decomposed into the single-atom response and the coherent addition of the extreme ultraviolet/soft X-ray (XUV) radiation during pulse propagation often called macroscopic effects. The three-step model [1,2] that includes tunnel ionization of electrons by the strong field, acceleration, and recombination with the parent atom provides an excellent understanding of the physics underlying the single-atom response. The tunnel ionization and acceleration are mostly determined by the driving electric field while recombination depends on the orbital structure of parent atom. The emitted XUV photon spectrum is directly related to the dipole moment due to the coherence between the returning electron wavepacket and the ionic ground state. The *ad hoc* factorization of the three-step model has been found to be very useful in the understanding of high-order harmonic (HH) spectra. The harmonic spectrum, *S*, can be described as follows [3-6]:

$$S(\omega) = \omega^4 |d(\omega)|^2 W(E) \propto \sigma^r(E) \cdot W(E), \quad (1)$$

where $\omega$ is photon energy of the emitted XUV radiation, $E$, the kinetic energy of returning electron with $E=\omega-I_p$ and $I_p$ being the ionization potential, $d$, transition dipole element in the length form, $W$, the electron wavepacket (EWP), and $\sigma^r$ is the photo-recombination cross-section (PRCS) that is related to photo-ionization cross-section (PICS). This shows that using accurate information on the EWP which is mainly determined by the driving laser field, we can extract the PRCS containing atomic or molecular orbital structure from the HH spectrum or vice versa. As studied by Le *et al.* [7] and known as quantitative rescattering (QRS) theory, the factorization with accurate PRCS can provide a quantitative correction to the single-atom HHG calculation based on strong-field approximation (SFA) which neglects the effect of atomic potential and assumes the plane wave of ionized electrons.

The HH spectrum, emitted by the rescattered EWP, shows the characteristic of long plateau and a sharp cutoff, while its shape is determined by the PRCS of the valence shell. However, the detailed spectral structure is also influenced by the propagation effects, such as the phase matching between XUV and driver pulses, the absorption of XUV pulses in the medium, and the plasma-induced driver pulse deformation [8]. The propagation effects change the macroscopic feature of coherent addition of XUV pulses and the ultimate efficiency of the HHG process as well as the HH spectral shape. In general, the HH spectroscopy is performed in the phase-matched regime with low medium density to avoid the propagation effect. However, it is not always possible to achieve the phase matching over broad harmonic bandwidth. Therefore, the quantitative analysis of the propagation effects always helps increasing the accuracy of HH spectroscopy. For example, the impact of propagation effects on the observed location of the Ar Cooper minimum has been studied very recently [9,10].

Experimentally, HHG has been mostly studied using 800-nm femtosecond Ti:sapphire laser amplifiers with or without external pulse compression [11]. Cutoff extension of HHG using long-wavelength drivers [12-16] has made seen rapid progress using drive laser technology based on infrared optical parametric amplification (OPA) [17] and optical parametric chirped-pulse amplification (OPCPA) [18]. As a result, recently it has been demonstrated that high-flux coherent XUV pulses with high photon energies beyond the keV region can be generated [19]. The unfavorable wavelength scaling of harmonic efficiency in the single-atom response (proportional to $\lambda^{-5}-\lambda^{-6}$) which is mainly due to the increased quantum diffusion can be compensated for to some extent by improved phase matching in HHG at high gas pressures exploiting the low absorption in helium for high photon energies [20]. Thus the efficiency study is important for developing high-flux coherent XUV sources.

Besides the development of high-flux XUV sources, long-wavelength-driven HHG is crucial for HH spectroscopy, because the extended cutoff and broad plateau give additional information on the single-atom response in the high photon energy region, revealing the atomic and molecular orbital structure more clearly. Recently, Shiner *et al.* [6] have observed the giant resonance in Xe, already known from earlier studies in physics of XUV region [21, 22], using a 1.8-μm driver and Vozzi *el al.* [23] have tomographically reconstructed the orbital structure of diatomic molecules using 1.3-1.7 μm drivers. On the other hand, the longer driver wavelength and lower efficiency makes the HHG process more sensitive to propagation effects due to the enhanced spectral coverage over which it is difficult to maintain phase matching. Therefore, propagation effects need to be addressed for accurate analysis of atomic and molecular structures in the HH spectroscopy. Recently, Jin *et al.* [5] theoretically studied how to retrieve PRCS from HHG under the presence of macroscopic effects using 3-dimensional (3D) propagation simulations.

In this paper, we experimentally and numerically investigate the spectral signature of the single-atom response and the propagation effect in HHG driven by a 2.1-μm source in Xe, Kr, and Ar gas jets. Section II describes the 2.1-μm kHz-repetition-rate OPCPA system and the experimental measurement of the HH spectra and the extraction of conversion efficiency. In section III, we discuss how to calculate the differential PRCS of the sub-shells in each gas using the published PICS data [6] and describe the numerical simulation of HHG using a 3D propagation code [8] and the three step model [20,24] combined with the differential PRCS, following the concept of the QRS theory [7]. In section IV, we present the simulated HH spectra and reproduce the experimental ones from Xe, Kr, and Ar. The PRCS curve of each gas is also extracted from the experimental and simulated spectra. Detailed comparison of experimental and simulated HH spectra with the PRCS curves enables to separate the atomic response from the propagation effects. We also discuss the Cooper minima in Kr and Ar and the possibility of the multi-electron effect involvement in Xe and Kr. Section 4 contains the conclusion.

## II. HHG experiment using an ultrabroadband 2.1-μm, kHz OPCPA system

### II.1 Picosecond pump lasers and 2.1 μm OPCPA setup

For the long-wavelength-driven HHG, we have developed a mJ-level 2.1-μm ultrabroadband OPCPA system operating at kHz repetition rate [18,25]. Picosecond pump laser technology is crucial for the energy scaling of ultrabroadband OPCPA systems. The operating parameters of the 2.1-μm OPCPA system and its pump lasers are summarized as follows. This OPCPA system is simultaneously pumped by two lasers operating at a kHz repetition rate: 1) ~12 ps, 4 mJ, 1047 nm Nd:YLF amplifier and 2) ~14 ps, 13 mJ, 1029 nm cryogenic Yb:YAG amplifier. Both pump lasers are seeded by one Ti:sapphire laser for achieving optical synchronization. The Nd:YLF chirped-pulse amplifier (CPA) pumps the first two OPCPA pre-amplifier stages. The required and used pump energy for the first two stages of the OPCPA was 1.9 mJ out of 4 mJ. The cryogenic Yb:YAG laser pumps the final OPCPA stage for power amplification.

Ultrafast cryogenic Yb:YAG laser technology has been proven to overcome the limitations of *ps* Nd:YLF laser technology in terms of both energy and average power. The second pump source was modified from a multi-ten-mJ, *ps* cryogenic Yb:YAG CPA system [26]. The seed from the Ti:sapphire oscillator is stretched by a chirped volume Bragg grating (CVBG) pair to ~560 ps with 0.7 nm of bandwidth at 1029 nm, and then amplified by a regenerative amplifier and a double-pass amplifier that are both based on cryogenically cooled Yb:YAG as gain medium. We limited the maximum energy to 20 mJ with a relatively large pump beam size (~3.6 mm diameter) at the double-pass amplifier to ensure damage-free long-term operation. The pulse is compressed to 14.2 ps using a multi-layer dielectric grating pair with a throughput efficiency of 75%. The available pump energy at the third OPCPA stage was 13 mJ following the optics for energy and optical delay control.

Figure 1 illustrates the schematic of the 2.1-μm OPCPA system and HHG setup. The passively carrier-envelope phase (CEP)-stable 2.1-μm seed pulses, produced by intra-pulse difference frequency generation (DFG) in MgO:PPLN, are stretched and amplified to 2.5 μJ in the first OPCPA stage based on a MgO:PPLN crystal. The pulses are then stretched to ~9 ps in full width at half maximum (FWHM) using an acousto-optic programmable dispersive filter (AOPDF). The second OPCPA stage (MgO:PPSLT crystal), pumped by ~1.4 mJ of energy, amplifies the pulses to 25 μJ. The pulses are further stretched to ~14 ps (FWHM) for the third OPCPA stage based on a β-barium borate (BBO) crystal. The pump intensity at the third stage was ~40 GW/cm$^2$ at 13 mJ of energy. After optimization of temporal and spatial overlaps and incidence angle of the pump beam into the BBO crystal, we obtained a maximum energy of 0.85 mJ with a conversion efficiency of 7.5% including the reflection loss of the pump beam on the uncoated BBO crystal. The amplified spectrum from the third OPCPA stage shows a spectral bandwidth of ~470 nm (FWHM) centered at 2.1 μm, supporting a transform-limited pulse duration of 24.5 fs (3.5 optical cycles). The pulse duration after compression using Brewster-angle fused silica (Suprasil 300) was measured to be 31.7 fs corresponding to ~4.5 optical cycles. The near-field output beam profile of the 2.1-μm pulse shows near Gaussian profile with an estimated M$^2$ of near ~1.3 [25].

**II.2 HHG setup and experimental results**

The compressed 2.1-μm output was delivered to the vacuum HHG chamber. In the HHG experiments, the typical energy after the third OPCPA stage and a telescope was 0.5−0.6 mJ. Due to the reflection loss at the silver mirrors (10 reflections) for beam delivery and at a Brewster-angle Suprasil 300 compressor (8 times transmission through glass surfaces) we had maximum ~0.4 mJ of compressed pulse energy available on target. The laser energy and beam size (~6 mm of diameter) was further controlled using a variable aperture which is also useful for improving the phase-matching condition of HHG, and the 2.1-μm beam was focused onto a gas jet using an $f$=200 mm or $f$=250 mm CaF$_2$ lens. Before we measure the HH spectrum, the HHG signal was detected with an Al-coated XUV photodiode (AXUV100, IRD) after a 500-nm-thick Al filter and then magnified using a low-noise electronic amplifier that significantly improves the detection sensitivity.

After maximizing the XUV photodiode signal by changing the aperture diameter and the gas jet position relative to the beam focus, we measured the HH spectra using an XUV spectrometer with a multi-channel plate (MCP) detector [27]. Generated XUV pulses were collected and focused at the XUV spectrometer using a toroidal mirror. The driving 2.1-μm pulse was blocked by an X-ray filter, for which we used a 500-nm-thick Al, 500-nm-thick Be, or 400-nm-thick Zr filter depending on the XUV range of interest. The signal from the calibrated XUV photodiode together with the measured HH spectra is used to compute the conversion efficiency of the HHG process from the experimental data.

The measured HH spectra from Xe, Kr, and Ar are shown in Figs. 2(a)-(c), where the estimated laser intensities at the focus are (0.7±0.1)x10$^{14}$ W/cm$^2$ for Xe, (1.1±0.1)x10$^{14}$ W/cm$^2$ for Kr and (1.7±0.2)x10$^{14}$ W/cm$^2$ for Ar. The estimated Gaussian beam waist at focus was in the range of 50–75 μm depending on gas species and focusing lens when the XUV photodiode signal was maximized by adjusting the aperture. The interaction length in the gas medium was ~2 mm and the gas pressure measured at the interaction region was ~50 mbar for Xe and Kr, and ~70 mbar for Ar. The gas pressure was limited by the maximum backing pressure (3 bars) that can be handled by our current pressure regulator. The Rayleigh range of the focused beam is more than 3.7 mm. The position of the gas jet relative to the focus was also scanned for maximizing the HHG efficiency in the Xe and Kr measurements. As a result, in Fig 2 the gas jet was placed slightly after the focus, which is a typical geometry for maximizing phase-matched short-trajectory high harmonics in the presence of Guoy phase [28].

Figure 2(a)-(c) clearly shows the cutoff extension from each gas compared to the conventional 800-nm-driven HHG. The cutoff energies in Xe, Kr, and Ar are at ~85 eV (~14 nm, ~149th harmonic), ~120 eV (~10 nm, ~211th harmonic) and ~160 eV (~7.8 nm, ~269th harmonic), respectively. The Al and Be edges are clearly shown in the spectra. The cutoff energy is typically <35 eV for Xe and <60 eV for Ar in the 800-nm-driven HHG experiments. The HHG peak efficiencies per harmonic near cutoff were measured to be ~1x10$^{-9}$ for Xe at ~70 eV, ~0.8x10$^{-9}$ for Kr at ~110 eV, ~2x10$^{-9}$ for Ar at ~130 eV, respectively. In terms of photon flux, the number of photons per second over 1% bandwidth at 130 eV for Ar is as high as 0.8x10$^8$. In our previous study [27], we obtained ~1x10$^{-8}$ efficiency at ~110 eV from He using an 800-nm driver in a loosely focused geometry. Despite the wavelength scaling of $\lambda^{-5}$–$\lambda^{-6}$, the measured efficiency using 2.1-μm driver did not significantly drop because of the different atomic properties between Kr and He. A multi-mJ, kHz 2.1-μm driver is expected to deliver 10$^9$–10$^{10}$ photons per second over 1% bandwidth in the phase-matched water-window soft X-ray HHG if high-pressure gas targets of Ne and He are used [8].

In view of HH spectroscopy, the HH spectra in Fig. 2 also contain some interesting features, such as the local minima and low-energy depletion. The detailed structure of the HHG spectra can be explained by the mixture of atomic response and propagation effect. We will explore the HH spectroscopy in the presence of macroscopic effects in the following sections.

### III. Single-atom response and propagation effects in the HH spectra

As indicated by Eq. (1), the spectral response of the PRCS can be studied by HH spectra if the EWP is well characterized. Since the HHG process is dominated by the electron dynamics in the valence shell, the atomic or molecular orbital in the valence shell can reconstructed by analyzing the HH spectra [3]. In the same context, HH spectroscopy is useful for probing multi-electron effects, which can significantly modify the HH spectrum generated from the valence shell electrons. Recently, the observations of multi-electron inelastic scattering dynamics between the returning electron, ionized from the valence shell, and the electrons in a sub-shell or inner-shell have been reported [6,29] using long-wavelength drivers. The multi-channel effect can be more easily observed in molecules [30] than in atoms because the ionization potential of the highest occupied molecular orbital (HOMO) is relatively close to the next-order orbital, such as HOMO-1 or HOMO-2.

On the other hand, the absorption of XUV pulses in the HHG medium itself and the phase matching between XUV pulse and driving laser pulse are the major macroscopic effects, strongly modifying the HH spectra as well. In this section, we discuss 1) the differential PRCS curves for valence shell and outermost sub-shells, 2) the 3D propagation simulation including absorption, phase mismatch, and plasma defocusing, and 3) the reconstruction of HH spectra using the given PRCS or extraction of PRCS from given experimental data. This procedure enables to analyze the possibility of multi-electron dynamics involvement in the HHG process by decomposing the macroscopic effects from the experimental HH spectra.

### III.1 Differential PRCS curves

Since the photo-recombination is the time-reversed process of photo-ionization, the PRCS can be retrieved from the photo-ionization cross-section (PICS) using the principle of "detailed balance" [31,32]. Since the ionization and recombination of electrons during HHG occurs predominantly along the laser polarization direction, we are interested in differential PICS (or PRCS) of the orbital relevant to the HHG process. Differential PICS for each orbital can be obtained from the quantum-mechanical calculations of the transition dipole moment between the EWP and the ionic state, where the modeling of the short-range potential is very critical for the accurate prediction of the wave function interference. The position sensitivity of the observed Cooper minimum of Ar is a good example. Here, following the approach in Refs. [4,6], we used the PICS ($\sigma^i$) curve and the angular asymmetry parameter ($\beta$) for each sub-shell published in Ref. [33] to obtain the differential PICS curve in Xe, Kr, and Ar. The total PICS curve, or the sum of the PICS curves for each orbital (partial PICS), shows a good agreement to more recent measurements on the total PICS of Xe, Kr, and Ar [34]. This confirms that the differential PICS calculated from Ref. [33] is reliable. The differential PICS is converted into the differential PRCS using detailed balance. In summary, the differential PRCS, $d\sigma^r/d\Omega$, in HHG, where the dipole is parallel to the polarization direction of the XUV photon and the driving field, can be calculated, in atomic units, by

$$\frac{d\sigma^r}{d\Omega} = \frac{\omega^2}{c^2 k^2} \frac{\sigma^i}{4\pi}[1+\beta(\omega)] = \frac{\omega^2}{2c^2(\omega-I_p)} \frac{\sigma^i}{4\pi}[1+\beta(\omega)], \qquad (2)$$

where $\sigma^r$ is the PRCS, $\Omega$ is the solid angle, $\omega$ is the XUV photon energy, $c$ is the speed of light, $k$ is the momentum of electron, $\sigma^i$ is the PICS, $\beta$ is the angular asymmetry parameter, and $I_p$ is the ionization potential of each sub-shell. As mentioned above, $\sigma^i$ and $\beta$ are taken from Ref. [33], and $\omega = I_p + E = I_p + k^2/2$. The calculated differential PRCS of each sub-shell in different gases will be presented and compared to the experimental HH spectra in section IV.

### III.2 3D propagation simulation of HHG
The 3D HHG simulation is a combination of three simulation steps: the propagation of the driver pulse, the single atom HHG at each propagation point, and the integration of emitted HH fields. Since the conversion efficiency of HHG is very small, we can ignore the influence of the HH field on the driver pulse and solve the propagation independently. The solution is the electric field distribution of the driver pulse at every spatial point as a function of time, with which we calculate the local electric dipole and the radiated HH field. Finally, these HH fields emitted from a volume element of the medium are linearly propagated through the rest of the medium and integrated at the end.

For propagation of the driver field, we use the propagation equation presented in Ref. [8], which takes into account diffraction, self-focusing by neutral media, plasma defocusing, and ionization loss. Dispersion and other nonlinear effects can be neglected because the medium under consideration is a low pressure short gas jet. Within the slowly evolving wave approximation, the propagation equation reads:

$$\frac{\partial E}{\partial z} = \frac{i}{2k} \nabla_\perp^2 E + i \frac{k}{2} n_2 \varepsilon_0 c |E|^2 E - \frac{1}{2c} \int_{-\infty}^{\tau} \omega_p^2 E d\tau' - \frac{I_p}{2c\varepsilon_0 \mathrm{Re}(E)^2} \frac{\partial \rho}{\partial \tau} E, \qquad (3)$$

where $E$ is the complex representation of the electric field; $z$ and $\tau$ are the propagation distance and the retarded time respectively; $k$ is the wave-vector at the carrier frequency; $\nabla_\perp^2$ is the transversal Laplace operator; $n_2$ is the nonlinear index of refraction; $\omega_p$ is the plasma frequency; $I_p$ is the ionization potential of the atom; $\rho$ is the number density of the ionized atoms. The ionization rate $\partial \rho / \partial \tau$ is given by the Ammosov-Delone-Krainov (ADK) formula [35]. Because of cylindrical symmetry, $E$ is a function of $z$, $\tau$, and the radial coordinate $r$. The diffraction in the radial direction is included with the aid of the quasi-discrete Hankel transform [36]. The propagation step along the $z$ direction is not limited by the propagation equation but by the phase difference and the highest harmonic order. When the phase of the driver pulse changes by 0.1rad, it results in a phase difference of 10.1rad for the 101st harmonic, which is far beyond the Nyquist criterion. Therefore, in our simulation, the propagation step is dynamic and ensures a phase change of the driver pulse less than 6.2 mrad, which is precise enough for sampling of a harmonic with 300 eV of photon energy. This propagation step is also small enough to accurately propagate the driver pulse.

### III.3 Three-step model and macroscopic HH spectrum
The single atom dipole response of HHG is calculated by the semi-classical three-step model. The HH dipole acceleration of a single atom can be written as [20]:

$$\ddot{x}(t) = \pi^{-1/2} e^{-i\pi/4} \omega_0^{3/2} (2I_p)^{1/4} \sum_n \frac{a(t_n')a(t)\sqrt{w(E(t_n'))}}{E(t_n')[\omega_0(t-t_n')/(2\pi)]^{3/2}} a_{rec} e^{-iS_n(t)}, \qquad (4)$$

where $\omega_0$ is the driver frequency; the summation over $n$ considers the possibility of multiple returns; $t'$ denotes the birth time of the ionized electron; $|a(t)|^2$ is the ground state probability; $w(E)$, as a function of the driver electric field, is the ionization rate given by the ADK formula; $a_{rec}$ is the recombination amplitude; $S_n(t)$ is the classical action. Here the acceleration gauge is employed to calculate the HH dipole acceleration [37], and the recombination amplitude is $a_{rec} = -\langle 0|\nabla V|\mathbf{k}\rangle$, where $|0\rangle$ is the ground state wave-function; $\nabla V$ is the gradient of the atomic potential; $|\mathbf{k}\rangle$ is the wave-function of a plane wave with momentum $\mathbf{k}$. It can be related to the conventional definition $\langle 0|\mathbf{x}|\mathbf{k}\rangle$ by:

$$a_{rec} = -\langle 0|\nabla V|\mathbf{k}\rangle = \left(I_p + \frac{\mathbf{k}^2}{2}\right)^2 \langle 0|\mathbf{x}|\mathbf{k}\rangle = \omega^2 d(\omega). \quad (5)$$

Because of quantum diffusion and phase matching constraints [38], the higher order returns of the ionized electron can be neglected, and only the first return and recombination is taken into account in our calculation. High-order returns also contribute only low photon energy part in HH spectra [39,40]. This eliminates in Eq. (4) the summation leaving only the latest trajectory. Transformed to the frequency domain and multiplied by appropriate pre-factors, the dipole acceleration can give us the single-atom HH spectrum. By setting the recombination amplitude $a_{rec}$ to be a constant (*i.e.*, unity), we can obtain the pure EWP spectrum as follows:

$$S(\omega) = \left|\int \ddot{x}(t) e^{i\omega t} dt\right|^2 = \omega^4 |d(\omega)|^2 W(E) = W(E) \equiv S_{flat}(\omega). \quad (6)$$

If combined with the pulse propagation effect, the calculated HH spectrum, $S_{flat}(\omega)$, directly represents the macroscopic EWP [5,41]. We integrated the emitted spectrum over all the radial position, but the spatial effect can be also investigated using our 3D propagation code.

The recombination amplitude and the actual HH spectrum can be calculated using the atomic structure represented by the differential PRCS [7].

$$\omega^4 |d(\omega)|^2 = \frac{c^3}{4\pi^2} \omega k \cdot \frac{d\sigma^r}{d\Omega} = \frac{\sqrt{2}c^3}{4\pi^2} \omega \sqrt{\omega - I_p} \cdot \frac{d\sigma^r}{d\Omega}. \quad (7)$$

Finally, we obtain the following relation for the HH spectrum:

$$S(\omega) \approx \omega \sqrt{\omega - I_p} \cdot \frac{d\sigma^r}{d\Omega} \cdot S_{flat}(\omega), \quad (8)$$

or equivalently

$$\frac{d\sigma^r}{d\Omega} \approx \frac{1}{\omega \sqrt{\omega - I_p}} \frac{S(\omega)}{S_{flat}(\omega)}. \quad (9)$$

Since this HH spectrum contains both the propagation ($S_{flat}(\omega)$) and atomic response ($d\sigma^r/d\Omega$), it can quantitatively reproduce the experimentally obtained data ($S(\omega)$) using Eq. (8). On the other hand, the differential PRCS ($d\sigma^r/d\Omega$) can be extracted from the experimental HH spectrum ($S(\omega)$) and the simulated macroscopic EWP ($S_{flat}(\omega)$) using Eq. (9).

### IV. Signature of single-atom response and propagation effects in HH spectra
In this section, we calculate the macroscopic EWPs in the HH spectrum generated with Xe, Kr, and Ar, and then reproduce the experimentally observed HH spectra using the differential PRCS of each gas computed from PICS. A detailed analysis of HH spectra enables to understand the signature of atomic response and propagation effects in the HH spectrum.

### IV.1 Analysis of HHG spectrum in Xe
We first studied the impact of propagation effects on the shape of the HH spectrum in Xe using the 3D propagation simulation that allows precise calculation of the macroscopic EWP ($S_{flat}(\omega)$) with a constant recombination amplitude. When the gas species and the laser beam profile are fixed, the pulse propagation effect is mostly determined by laser intensity, focusing geometry, gas pressure, and medium length. In the simulation, we set all the parameters to be close to the experimental values. Since the gas jet position ($z$) relative to the focus is one of the critical macroscopic parameters for efficient HHG, we plotted the HH spectra depending on $z$. Figure 3(a) illustrates the definition of sign of gas jet position ($z$), where we set $z$ to be negative when the laser beam focus is before the gas nozzle. The XUV absorption by the gas medium itself is one of the major macroscopic effects affecting the spectral shape and is represented by the Xe transmission curve in Fig. 3(b). A medium length of 2 mm and a pressure of 50 mbar are used in the simulation as the experimental conditions in Fig. 2(a).

Figure 3(c)-(e) shows the calculated spectra of macroscopic EWP depending on the $z$ position in terms of photon energy ($S_{flat}(\omega)$ in Eq. (6)) after the propagation of pulses. The focused intensity in this calculation is $0.8 \times 10^{14}$ W/cm$^2$, which is within the estimated experimental intensity. When the laser focus is placed 2 mm after the gas jet

(z=+2 mm) or the gas jet is placed at 2 mm before the focus, the cutoff energy is extended only up to 60 eV and the absorption in the low photon energy range is weak, as shown in Fig. 3(c). This means that the high harmonics are mostly generated at the end of the jet, experiencing low absorption, and the effective peak intensity of the pulse is low. Figure 3(d) is the EWP spectrum calculated when the gas jet is at the focus (z=0 mm). The cutoff energy of ~90 eV due to higher peak intensity and the low-energy (<30 eV) absorption are observed. Figure 3(e) is for the case that the gas jet is placed at 2 mm after the focus (z=-2 mm). Since the harmonics are generated at the beginning of the gas jet due to the focal position, the absorption at both the cutoff and low-energy part is very strong, as expected by the Xe transmission curve of Fig. 3(b). As a result, the cutoff is reduced to ~75 eV. The absorption region is marked as yellow shade. It is expected that the cutoff will be extended to beyond ~90 eV of Fig. 3(d) if the absorption is not strong at the cutoff, as will be discussed in Kr and Ar cases.

To reproduce the experimental HH spectrum in Xe with the simulation, we chose an EWP spectrum at a proper $z$ position close to the experimental condition and calculate the HH spectrum $S(\omega)$ using Eq. (8). The condition z=0 mm is not easy to clearly define by using the beam profile in the experiment because the imaging of the focused 2.1-µm beam is not trivial. Instead, we used the position of plasma generated in air at low laser intensity. The position of the gas jet for the experimental HH spectrum shown in Fig. 2(a) was slightly after the focus within the Rayleigh range. The experimental error in locating the focus position is estimated to be ~1 mm. After comparing the experimental HH spectra with the simulated HH spectra at z=0, -1, and -2 mm, we found the best fit for z=-1 mm.

Figure 4(a) shows the simulated macroscopic EWP ($S_{flat}(\omega)$) at z=-1 mm and Fig. 4(b) is the corresponding HH spectrum calculated using Eq. (8), where the differential PRCS curve of the valence 5p shell of Xe was obtained from Ref. [33] and Eq. (2). Figure 4(c) is the experimental HH spectrum in Xe whose spectral response was corrected from Fig. 2(a) for the X-ray filter transmission and detector sensitivity. Both calculated and measured spectra are compared with the 5p shell PRCS for indentifying the imprinted atomic response. The calculated HH spectrum quantitatively reproduces the cutoff energy (~90 eV) and low-energy absorption (<30 eV) of the experimentally measured HH spectrum. The experimental and calculated HH spectra of Figs. 4(b) and (c) generally follow the tendency of PRCS of 5p valence shell, as expected.

As an alternative way of analysis to the reproduction of experimental HH spectra, in Fig. 4(d), we extracted the PRCS curve (cross dots) of Xe from Figs. 4(a) and (c) using Eq. (9) and compare it with the differential PRCS curve of 5p shell (red solid line) from published data of Fig. 4(b). The macroscopic effect is cancelled out in the extracted PRCS curve, showing that the precise simulation of macroscopic EWPs enables to perform the experimental HH spectroscopy in the presence of propagation effects. Morishita et al. [42] theoretically proposed a similar method of obtaining PRCS curves using time-dependent Schrödinger equation and EWP in the absence of macroscopic effects. We extend this concept to the macroscopic EWPs [4]. In Fig. 4(d), for reference, the differential PRCS curves for the next outermost sub-shells, 5s and 4d, are plotted as blue dashed and magenta dotted lines, respectively. The ionization potentials for 5p, 4d, and 5s sub-shells are 12, 68, and 28 eV, respectively. Since the 5p shell is the valence shell, where the electron is born and tunnel ionized during the HHG process, the ionized electron is expected to recombine into the 5p shell. In contrast, the giant resonance at ~100 eV reported by Shiner et al. [6] is attributed to the indirect 4d shell contribution in the recombination process. It originates from the Coulomb-induced inelastic scattering between the ionized electron from 5p shell and the bound electron in 4d shell, which forces a 4d electron to make a transition to the 5p hole. Our measurement in Fig. 4(b) does not show the giant resonance because the observed cutoff is not extended enough to see the effect of the 4d shell. The laser intensity was high enough to reach >100 eV at the cutoff, but multiple measurements confirmed that the HH signal quickly dropped at a higher intensity and the observed cutoff energy was not extended to higher than 90 eV. The simulated spectrum of Fig. 4(b) along with the Xe absorption curve in Fig. 3(b) clearly indicates that this is due to the absorption at the cutoff region. This suggests that the extended cutoff to >100 eV could be observed with much lower pressure or shorter medium length at the cost of HH efficiency. In the current experiment, the efficiency dramatically dropped close to the detection limit of our XUV spectrometer at a lower pressure. Therefore, a higher dynamic range of the spectrometer or higher driving energy is necessary for the HH spectroscopy in a single-atom regime. It should be noted that the absorption length is inversely proportional to the total PICS. Therefore, the atomic property of the gas is important for both the single-atom and propagation effects in the HHG process in terms of differential PRCS and total PICS, respectively.

It is interesting to note in Fig. 4(d) that there is quantitative discrepancy found in the range of 60–80 eV between the extracted PRCS and the 5p shell PRCS curve. The former one has minor offset from the latter in this range. One possible way of explaining this offset is the contribution of the electron recombination into the 5*s* shell, which has a few times larger PRCS in this range than the 5*p* shell. If the inelastic scattering is considered, the 5*s* shell contribution modifies the effective PRCS that is proportional to both the Coulomb interaction term and the PRCS of the 5*s* shell. The Coulomb interaction is inversely proportional to the ionization energy difference ($\Delta I_p$) between 5*s* and 5*p* shells (16 eV), indicating that the electrons in closer shells experience larger Coulomb interaction. In fact, the 5*s* shell is much closer to the 5*p* shell than the 4*d* shell relevant to the giant resonance and $\Delta I_p$ between 5*p* and 4*d* shells is as large as 56 eV. Therefore, despite the relatively small PRCS of 5*s* shell compared to that of the 4*d* shell, the minor spectral enhancement seems feasible from the electron recombination into 5*s* shell via Coulomb interaction. The nontrivial quantum mechanical calculation of the Coulomb interaction term [6] is necessary to quantify the indirect 4*s* shell contribution. In general, recent theoretical studies of the multi-channel effect [43] on HHG in atoms and molecules will help to quantitatively understand the multi-electron dynamics of HHG.

### IV.2 Analysis of HHG spectrum in Kr

The same analysis is applied to the HH spectrum from Kr. We carried out 3D propagation simulations for 2-mm-long Kr gas jet at 50 mbar to investigate the spectral shaping due to the macroscopic effects. Figure 5(a) is the XUV transmission curve of the Kr gas medium and Figures 5(b)-(d) show the simulated macroscopic EWP spectra ($S_{flat}(\omega)$) for z=2, 0, and -2 mm, respectively, where the laser intensity at the focus is $1 \times 10^{14}$ W/cm$^2$. The dependence of the EWP spectra on the gas jet position can be explained in a similar way as for Xe in the previous section. One main difference is the extension of cutoff for z=-2 mm (~130 eV) compared to that for z=0 mm. We observe that the cutoff energy and the center of phase matching move towards higher photon energy as the gas jet move along the laser pulse through the focus. The cutoff is not limited by the absorption of Kr unlike for Xe, as shown by the transmission curve of Fig. 5(a).

Knowing the differential PRCS of the Kr valence shell and the macroscopic EWP, we reproduced the experimental HH spectrum by use of Eq. (8). In the experiment the gas jet was also placed slightly after the focus. Figures 6(a) and (b) show the simulated macroscopic EWP ($S_{flat}(\omega)$) at z=-1 mm and the corresponding HH spectrum considering the PRCS of the 4*p* valence shell, respectively. Figure 6(c) is the experimental HH spectrum whose spectral response is corrected from Fig. 2(b). For comparison, the differential PRCS of the 4*p* valence shell is also plotted in Figs. 6(b) and (c). The simulated spectrum reproduces not only the cutoff energy at ~l30 eV and the low-energy (<45 eV) absorption, but also the spectral enhancement in the range 60−100 eV in comparison with the PRCS of the 4*p* shell. As is well known for Ar, the Cooper minimum [44] in Kr is attributed to the interference between *s*- and *d*-waves in the continuum state during the recombination into the *p* shell [45]. The Cooper minimum of the 4*p* shell at ~80 eV of the PRCS curve, which is the main signature of the atomic response, is observed in both simulated and experimental spectra. However, due to the propagation effect, the observed position in the HH spectrum is shifted to the range of 60−70 eV in the experiment and 70−80 eV in the simulation.

As an alternative approach, we extracted the PRCS curve of Kr from Figs. 6(a) and (c) using Eq. (9), as we did with Xe. Figure 6(d) shows the extracted PRCS curve (cross dots) together with the differential PRCS curves of two outermost shells, 4*p* (red solid line) and 4*s* (blue dashed line) shells. The ionization energies for the 4*p* and 4*s* shells are 14 and 26 eV, respectively. The extracted PRCS fits well to the computed differential PRCS of 4*p* shell and the shift of the Cooper minimum is less dramatic than in the experimental spectrum because the macroscopic effect is mostly removed in the extraction process. However, there is still a small offset in the range of 70−100 eV compared to the PRCS of the 4*p* shell. Similar to the Xe case, it is interesting that the PRCS of the 4*s* shell increases from 45 to 120 eV. If the Coulomb interaction is strong enough, involvement of the 4*s* shell may explain the shallower minimum of our derived total PRCS compared to the published differential PRCS of the 4*p* shell. This can also lead to the argument of the indirect contribution of the 4*s* shell via Coulomb interaction like the case of Xe 5*s* shell, which still requires further theoretical and experimental investigations. Even though its origin is not yet conclusive, the offset of the extracted PRCS in the range of 70−100 eV helps to explain the dramatic shift of the Cooper minimum observed in our experimental HH spectrum in addition to the macroscopic effect.

### IV.3 Analysis of HHG spectrum in Ar

The last gas medium we studied using HHG is Ar. To quantitatively investigate the propagation effect, we again performed a 3D propagation simulation with a 2-mm-long Ar jet at 70 mbar of pressure. Figure 7(a) is the XUV transmission in Ar and Figures 7(b)-(d) show the calculated macroscopic EWP of HHG in Ar for z=2, 0, -2 mm, respectively, where the laser intensity is $1.5 \times 10^{14}$ W/cm$^2$. The tendency with the gas jet position scan is basically the same as in Kr. Phase matching is shifted to higher photon energy and the low-energy absorption is stronger when the gas jet is placed after the focus (z=-2 mm). The cutoff is not limited by absorption of Ar.

The Cooper minimum of Ar at 45−55 eV has been well known and studied extensively with PICS [34] and HHG [46,9,10]. The position of calculated Cooper minimum depends on the short-range potential model [7] and the experimental observation also depends on the experimental conditions. The PRCS curves for the outermost sub-shells, 3$p$ and 3$s$, are plotted for comparison in Fig. 8(d), where the ionization energy is 16 and 28 eV, respectively. The minima at ~53 eV for the 3$p$ shell and ~43 eV for the 4$s$ shell obtained from Ref. [33] are consistent with the measured minimum at ~47 eV for the total PICS in Ref. [34] that is the mixture of 3$p$ and 3$s$ shells. The Cooper minimum at ~53 eV has been observed in a very recent HH spectroscopy experiment using a 1.8-μm driver by Higuet *et al.* [9].

We reproduced the HH spectrum from the PRCS of the Ar 3$p$ valence shell and the simulated EWP using Eq. (8). The PRCS curve for >110 eV is linearly extrapolated as an approximation from the data of Ref. [33] because the total PICS in Ref. [47] shows a linear decrease in this range. The asymmetry parameter is constant at 2 for the 4$s$ shell while it gradually increases from 1.5 to 2 for the 4$p$ shell, indicating that there is no oscillation or resonance of PRCS in the extrapolated range. Figure 8(a) shows the HH spectrum at z=0 mm calculated from Fig. 7(c), which clearly reveals the Cooper minimum at ~53 eV, as expected by the PRCS of the 3$p$ shell. In contrast, the HH spectrum at z=-2 mm of Fig. 8(b) calculated from Fig. 7(d), which is close to our experimental condition, does not show a clear minimum in the overall HH spectral structure. The phase-matched region, shifted to the high photon energy, and the low-energy absorption wash out the structure of a local minimum. This is consistent with the experimental HH spectrum in Fig. 8(c), where the spectral sensitivity of this figure is corrected from Fig. 2(c) using the detector sensitivity and X-ray filter transmission. The signal below ~55 eV was measured to be in the noise level and thus the Cooper minimum at ~53 eV is not clearly observed. This result shows that the macroscopic effects can make it difficult to observe the Cooper minimum in such experiments as optimized for high conversion efficiencies.

We also extracted the PRCS of Ar from Figs. 8(c) and 7(d) using Eq. (9), as shown by cross dots in Figure 8(d). The differential PRCS curves of 4$p$ (red solid line) and 4$s$ (blue dashed line) shells are also presented for comparison, where the extrapolated parts (>110 eV) are represented by thin dotted lines. The extracted PRCS is in good agreement with the 4$p$ PRCS curve in the range of 55−110 eV. The discrepancy for >100 eV is attributed to the cutoff behavior of the experimental HH spectrum as well as the approximation in the extrapolation of the 4$p$ PRCS curve. The indirect contribution of the 3$s$ shell does not need to be considered because its PRCS is already much lower than the 3$p$ PRCS. Even though the Cooper minimum was not revealed, the accuracy of extracted PRCS in the range of 55−110 eV demonstrates that the HH spectroscopy is still possible under the strong macroscopic effects.

**V. Conclusions**
We studied the atomic response and pulse propagation effects in HHG with Xe, Kr, and Ar driven by a 2.1-μm femtosecond OPCPA system. The cutoff extension was demonstrated and single harmonic conversion efficiencies were measured in the range of ~$10^{-9}$ near the observed cutoff energy. The extended cutoff in the long-wavelength driven HHG has revealed the spectral shaping of high-harmonics due to both the PRCS and the pulse propagation, which are the two main factors that determine the conversion efficiency besides the driving wavelength. Following the concept of QRS theory, we numerically reproduced the experimental HH spectra using a modified strong-field approximation and a 3D pulse propagation simulation. By extracting PRCS curves from the experimentally measured HH spectra and simulated macroscopic EWP, we showed that the HH spectroscopy is still feasible in the presence of propagation effects because the atomic response can be quantitatively separated from the propagation effects with the help of precise calculation of the macroscopic EWP.

Since the calculation of macroscopic EWP is mostly determined by the driving laser field, the ionization potential, and macroscopic parameters, it can be extended to the HHG in molecules. Therefore, our approach makes the HH spectroscopy more powerful tool for analyzing the atomic and molecular structure under the propagation effects. More systematic comparison of experimental and simulated HH spectra along with an extensive parameter scan will help to increase the precision of this method.

**Acknowledgement**
This work was supported by AFOSR FA9550-09-1-0212, FA9550-10-1-0063, and FA9550-12-1-0080, and the Center for Free-Electron Laser Science, DESY.

# Figures

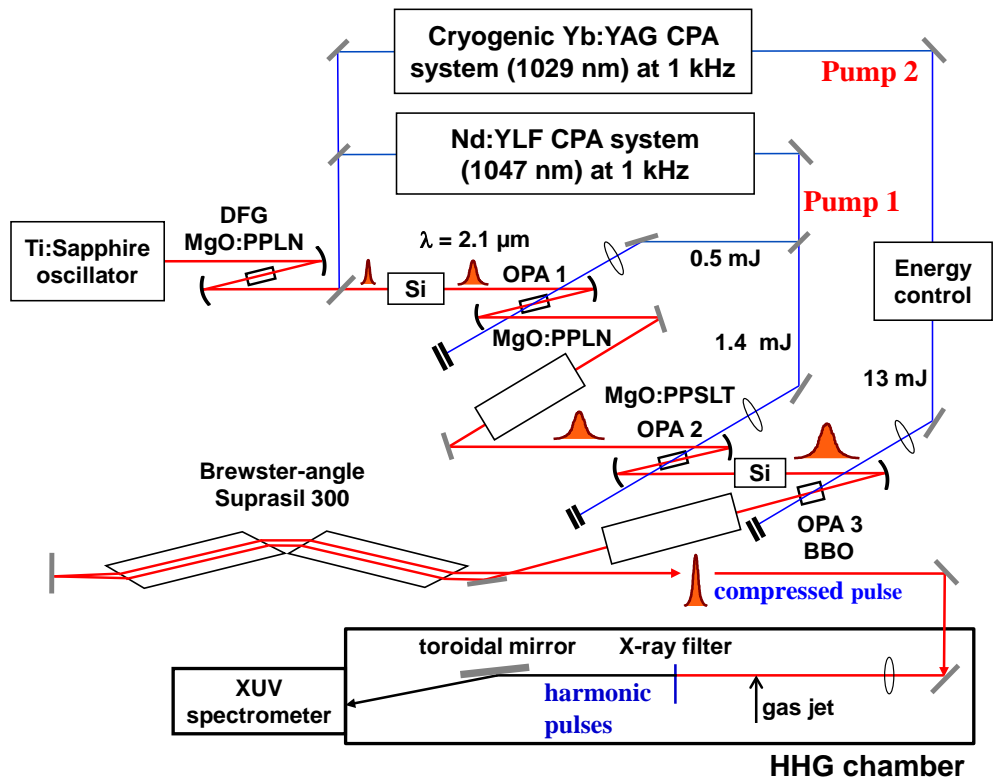

Fig. 1 Optical layout of the ultrabroadband CEP-stable 2.1-μm 3-stage OPCPA system and HHG setup. Acronyms are described in the text.

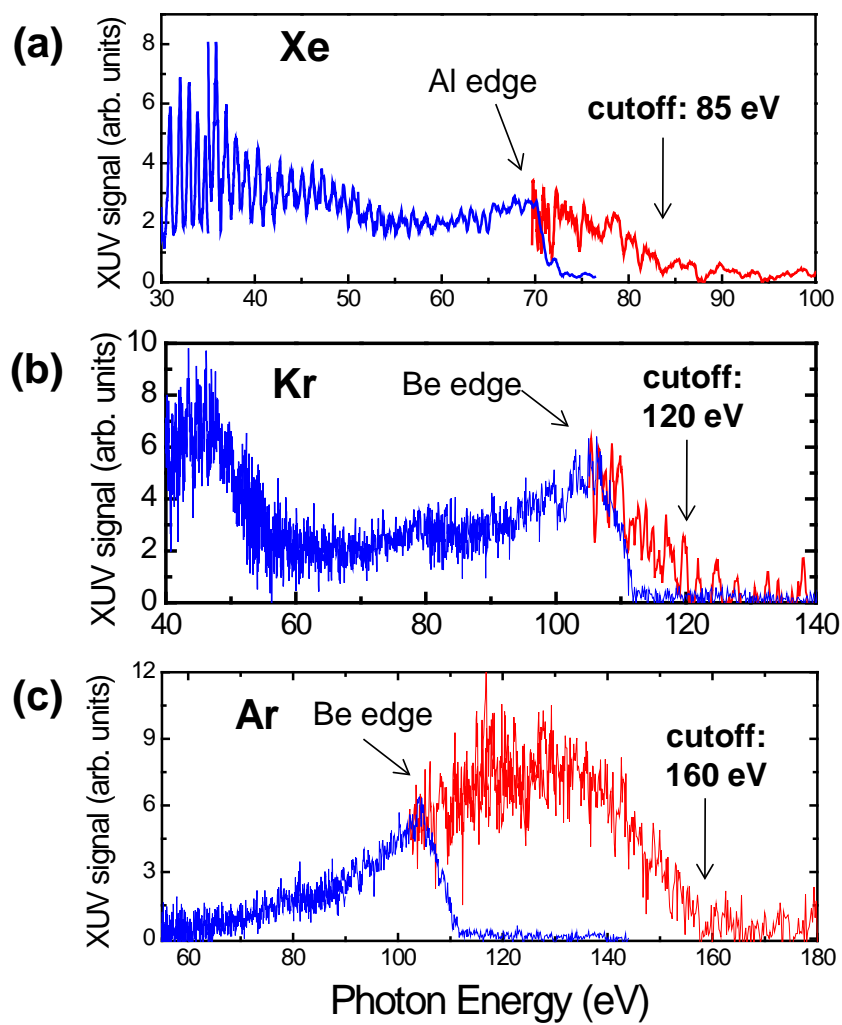

Fig. 2 (Color online) Experimentally measured HH spectra from Xe (a), Kr (b), and Ar (c). The blue and red curves show the measurements with different X-ray filters.

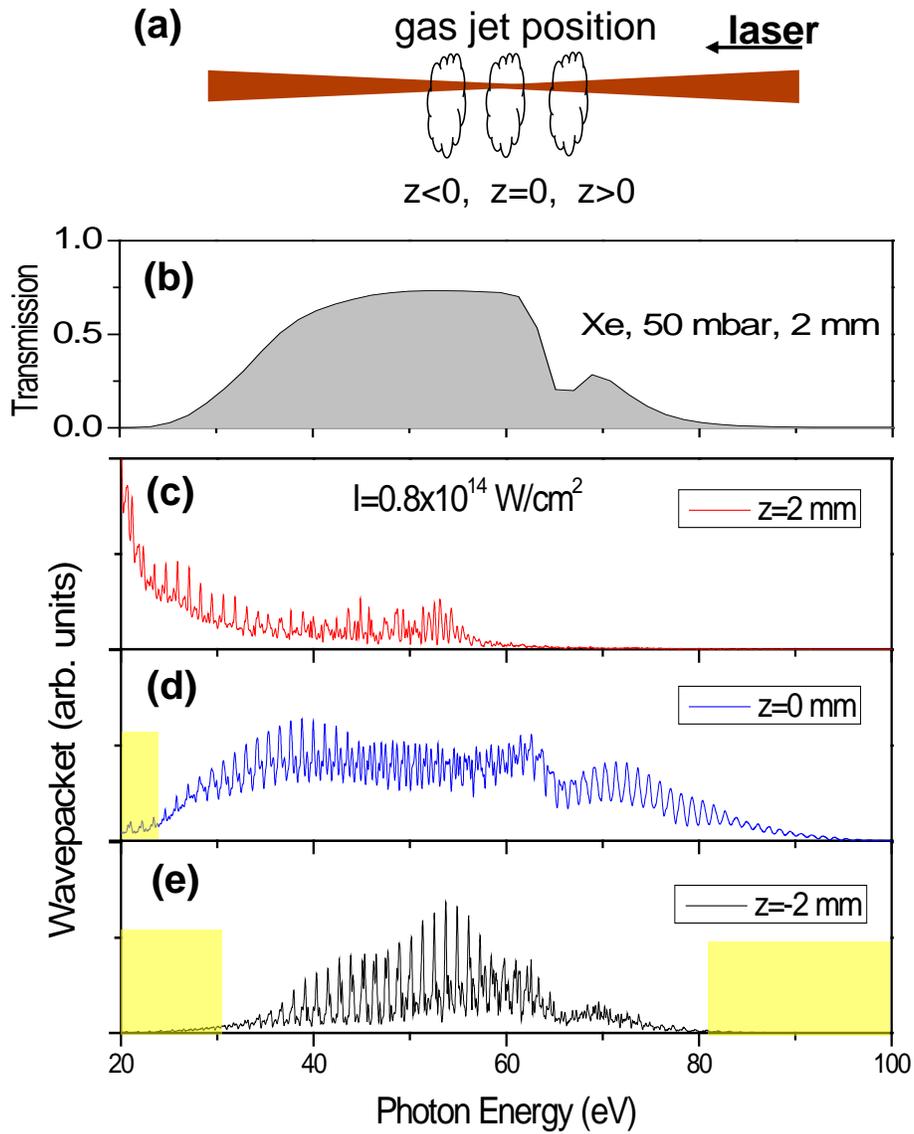

Fig. 3 (Color online) Calculated macroscopic EWP ($S_{flat}(\omega)$) depending on the Xe gas jet position. Definition of the sign of $z$ (a), XUV transmission in 2-mm-long Xe medium with 50 mbar (b), and EWP spectra for z=+2 mm (c), z=0 (d), and z=-2 mm (e). Yellow area represents the region affected by strong XUV absorption.

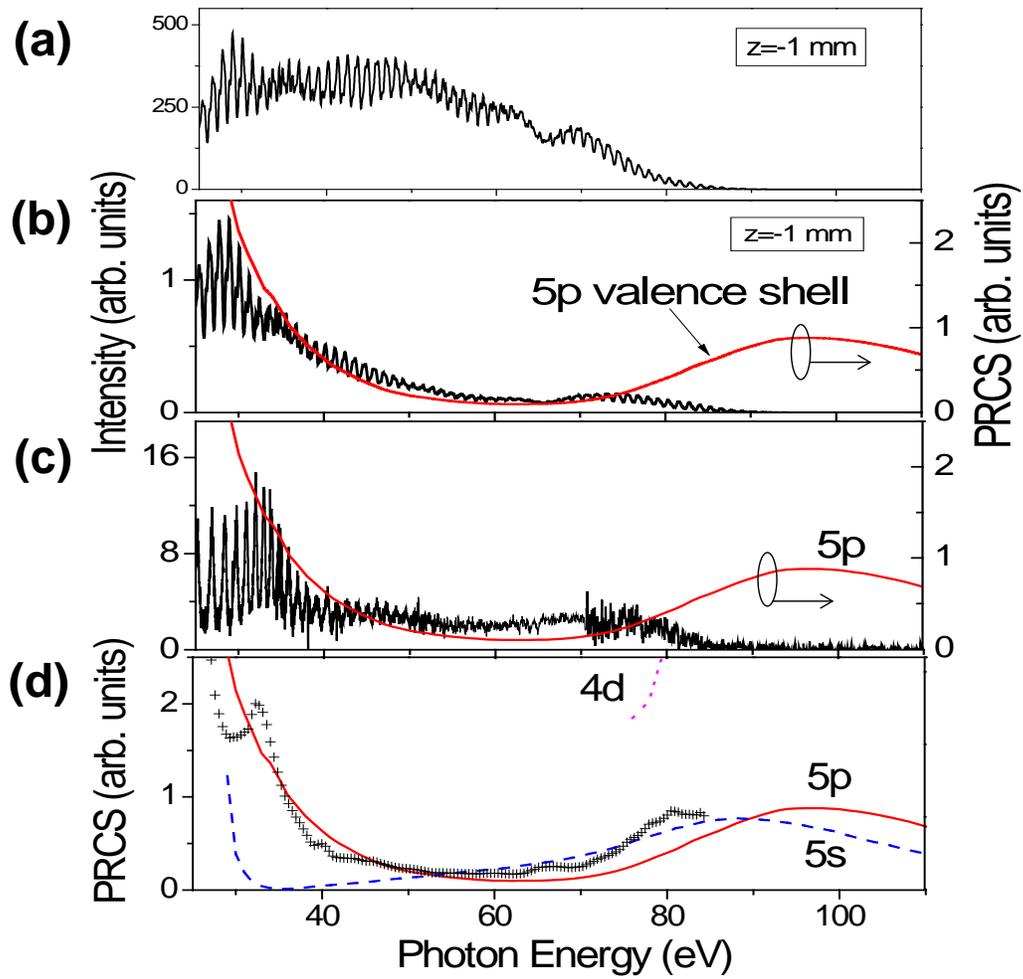

Fig. 4 (Color online) Reproduction of the experimental HH spectrum in Xe using simulation and comparison of PRCS curves. Simulated macroscopic EWP ($S_{flat}(\omega)$) at z=-1 mm (a), calculated spectrum with PRCS curve of 5$p$ shell (b), experimental HH spectrum corrected for spectral response of detector and X-ray filter (c), and the extracted PRCS (cross dots) in comparison with the PRCS curves of 5$p$, 5$s$, and 4$d$ shells, represented by red solid line, blue dashed line, and magenta dotted line, respectively (d).

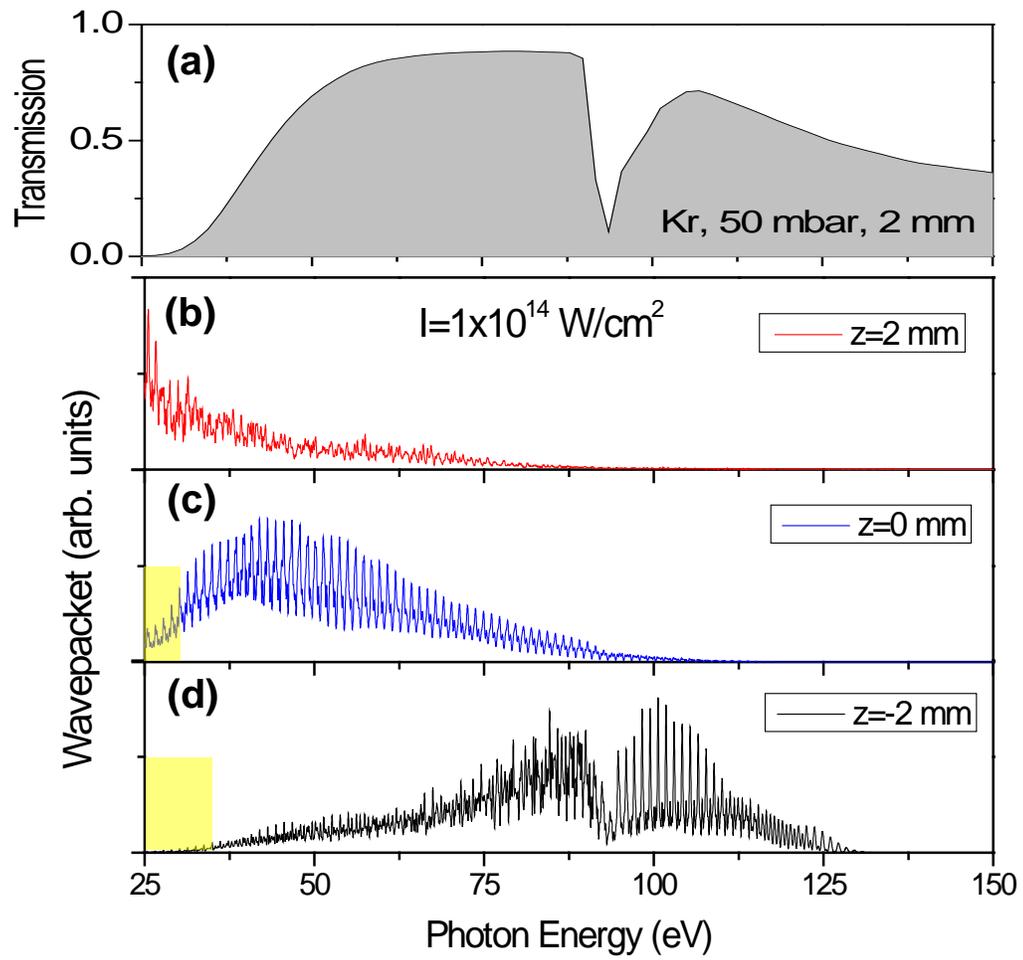

Fig. 5 (Color online) Calculated macroscopic EWP ($S_{flat}(\omega)$) depending on the Kr gas jet position. XUV transmission in 2-mm-long Kr medium with 50 mbar (a), and calculated EWP spectra for z=+2 mm (b), z=0 (c), and z=-2 mm (d). Yellow area represents the region affected by strong XUV absorption.

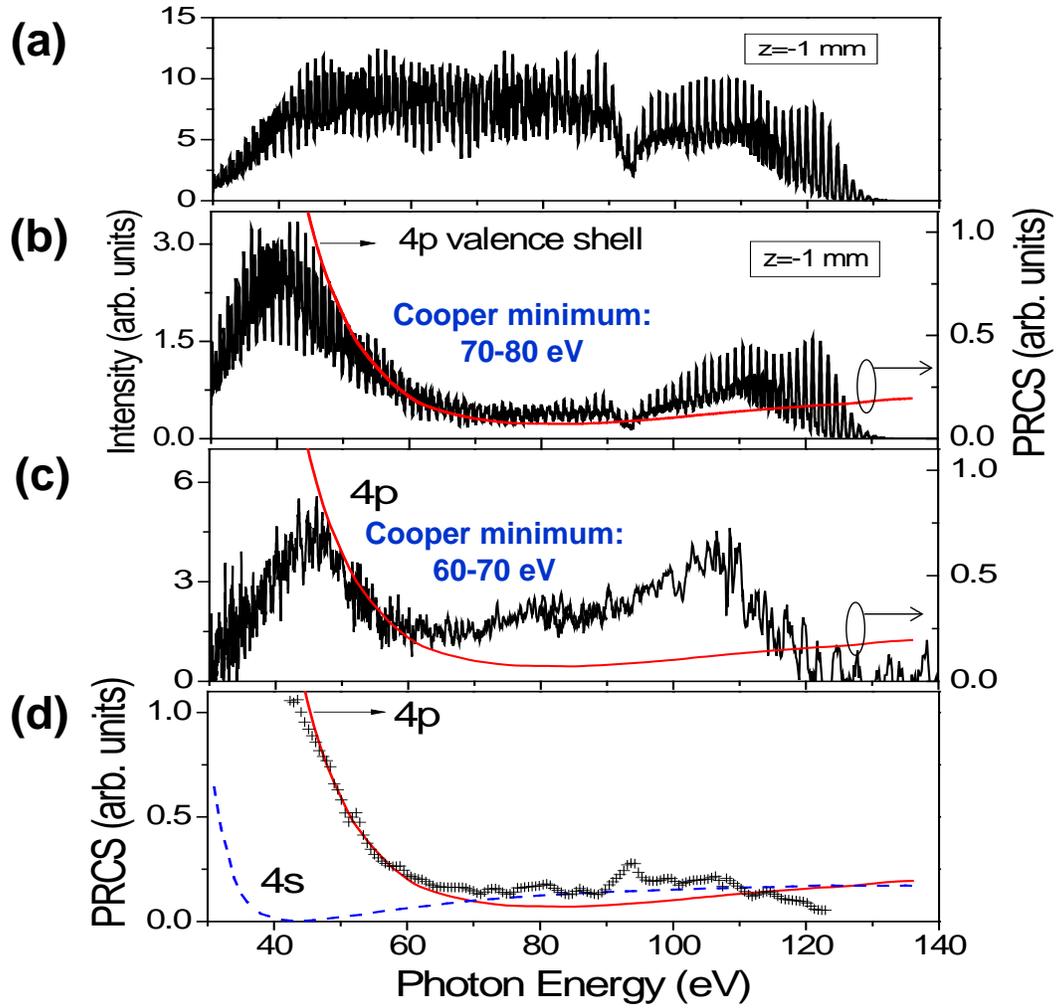

Fig. 6 (Color online) Reproduction of the experimental HH spectrum in Kr using simulation and comparison of PRCS curves. Simulated macroscopic EWP ($S_{flat}(\omega)$) at z=-1 mm (a), calculated spectrum with PRCS curve of 4$p$ shell (b), experimental HH spectrum with the correction of spectral response of detector and X-ray filter (c), and the extracted PRCS (cross dots) compared with PRCS curves of 4$p$ and 4$s$ shells, represented by red solid line and blue dashed line, respectively (d).

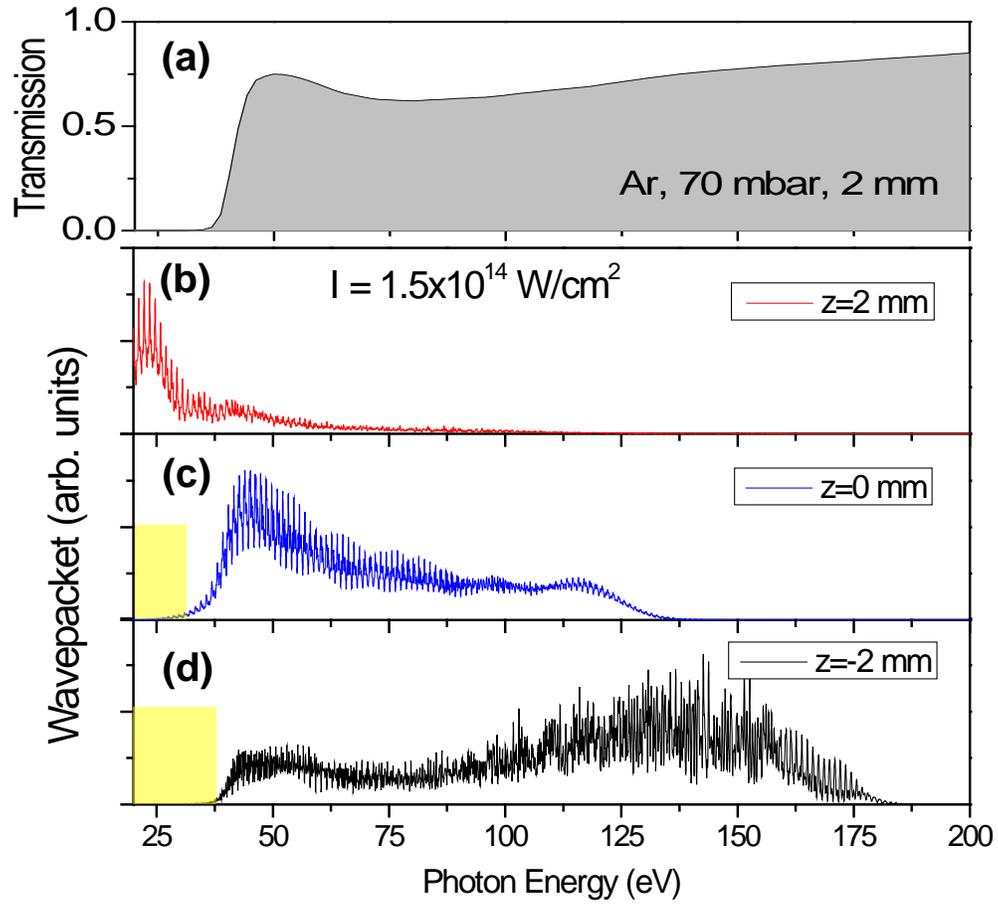

Fig. 7 (Color online) Calculated macroscopic EWP ($S_{flat}(\omega)$) depending on the Ar gas jet position. XUV transmission in 2-mm-long Ar medium with 70 mbar (a), and calculated EWP spectra for z=+2 mm (b), z=0 (c), and z=-2 mm (d). Yellow area represents the region affected by strong XUV absorption.

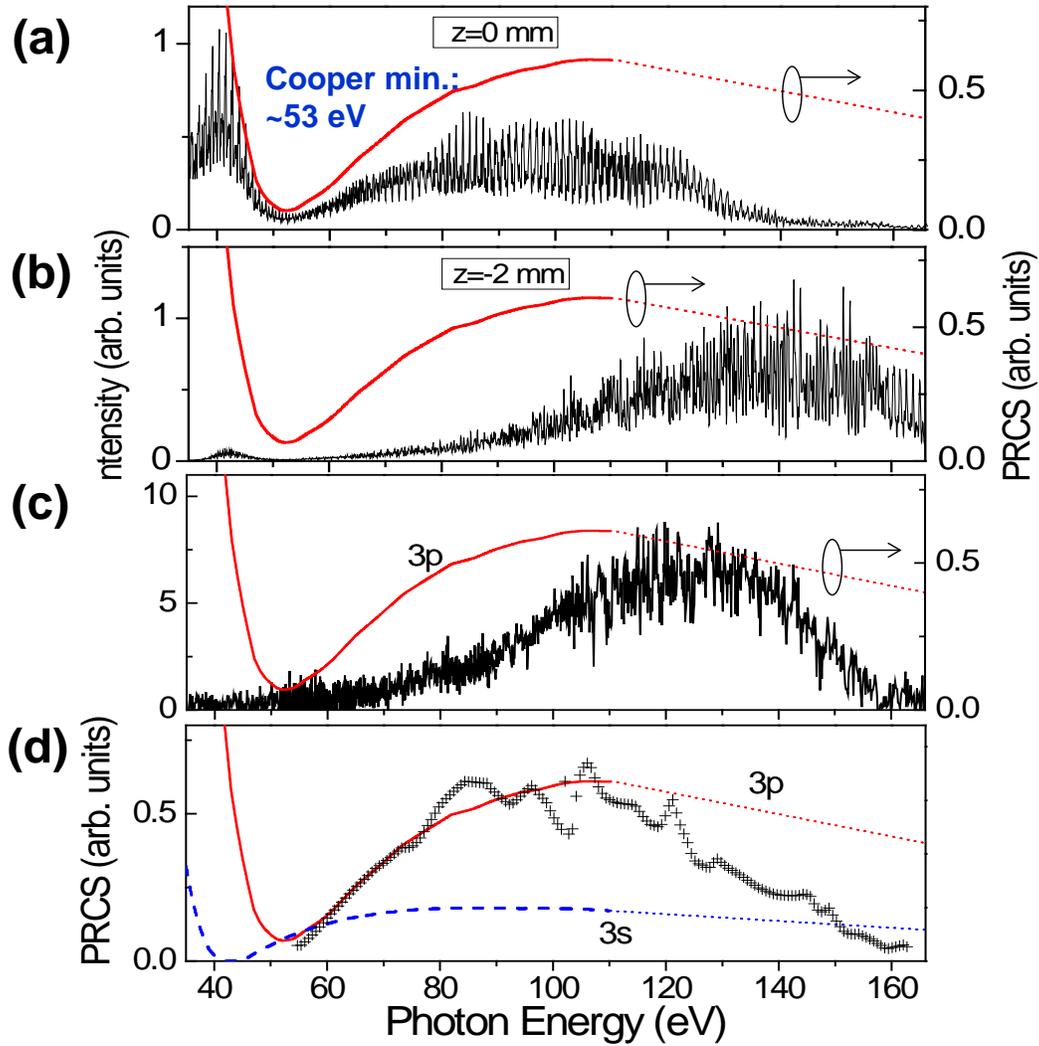

Fig. 8 Comparison of calculated HH spectra at z=0 (a) and -2 mm (b) in Ar with the experimentally measured spectrum (c), and the comparision of PRCS curves. (c) is corrected for the spectral response of detector and X-ray filter. The PRCS curves of 3$p$ and 3$s$ shells in (d) are represented the red solid line and blue dotted line, respectively, while the extracted PRCS curve is by cross dots. The dotted lines for >110 eV in all figures are linearly extrapolated [33, 47].